\documentclass[english]{article}

\usepackage{geometry}
\usepackage{graphicx}
\usepackage{amssymb}
\usepackage{epstopdf}
\usepackage{babel}
\usepackage{inputenc}
\usepackage{epsfig}

\begin{document}

\begin{center}
{\Large {\bf Fragmentation cross sections of {\boldmath Fe$^{26+}$, 
Si$^{14+}$ and C$^{6+}$} ions of {\boldmath $0.3 \div 10$} A GeV on 
polyethylene, CR39 and aluminum targets }} \\
\vspace{0.8cm}
S. Cecchini$^1$,
T. Chiarusi$^1$,
G. Giacomelli$^1$, 
M. Giorgini$^1$,  
A. Kumar$^{1,4}$,     
G. Mandrioli$^1$,   
S. Manzoor$^{1,2,3}$,  
A. R. Margiotta$^1$,  
E. Medinaceli$^1$,  
L. Patrizii$^1$,  
V. Popa$^{1,5}$,  
I. E. Qureshi$^{2,3}$,
G. Sirri$^1$, 
M. Spurio$^1$ and 
V. Togo$^1$ 

\par~\par

{\it  1. Phys. Dept. of the University of Bologna and INFN, Sezione di 
Bologna, Viale C. Berti Pichat 6/2, I-40127 Bologna, Italy \\ 
2. PD, PINSTECH, P.O. Nilore, Islamabad, Pakistan \\
3. COMSATS Institute of Information Technology 30, H/8-1, Islamabad, Pakistan\\
4. Dept. Of Physics, Sant Longowal Institute of Eng. and Tech., Longowal 
148 106, India \\
5. Institute of Space Sciences, Bucharest R-077125, Romania} 

\par~\par

\end{center}

\vspace{0.5cm}

{\bf Abstract.} {\normalsize We present new measurements of the total and 
partial fragmentation cross sections in the energy range $0.3 \div 10$ A GeV 
of $^{56}$Fe, $^{28}$Si and $^{12}$C beams on polyethylene, CR39 and 
aluminum targets. The exposures were made at BNL, USA and HIMAC, Japan. The 
CR39 nuclear track detectors were used to identify the incident and survived 
beams and their fragments. The total fragmentation cross sections for all 
targets are almost energy independent while they depend on the target 
mass. The measured partial fragmentation cross sections are also discussed.
}
\section{Introduction}
The interaction and propagation of intermediate and high energy heavy ions in 
matter is a subject of interest in the fields of astrophysics, radio-biology 
and radiation protection \cite{1}. An accurate description of the 
fragmentation of heavy ions is important to understand the effects of 
the high $Z$ component of Cosmic Rays (CRs) on humans in space \cite{2} and 
for shielding in space and in accelerator environments.  
 More recently the interaction and transport of light energetic ions in 
tissue-like matter became of particular interest in medicine and for 
hadron therapy of cancer \cite{4}. \par

When a heavy ion impinges on a target, it undergoes fragmentation processes 
depending on the impact parameter between the colliding nuclei. 
The target fragments carry little momentum. At high energies, the projectile 
fragments travel at nearly the same velocity 
as the beam ions and have only a small deflection. \par

The availability of heavy ion beams at the CERN SPS, at BNL (USA) and at 
the HIMAC (Japan) facilities made possible to investigate the projectile 
fragmentation on different targets and for different projectile energies. 
Several authors [4-10] 
have successfully used 
Nuclear Track Detectors (NTD's) for systematic measurements of nuclear 
fragmentation cross sections. \par

The present study is focused on Fe, Si and C ion interactions in CH$_2$, CR39 
$(C_{12}H_{18}O_7)_n$ and Al targets. We used CR39 detectors, which 
are sensitive for a wide range 
of charges down to $Z = 6e$ in the relativistic energy 
region \cite{11, 12}. NTD's have been used to search for exotic particles 
like Magnetic Monopoles and Nuclearites \cite{25,27}, to study cosmic ray 
composition \cite{26} and for environmental studies \cite{28}. 


\section{Experimental Procedure}
\begin{figure}[ht]
\centering
{\centering\resizebox*{!}{4.0 cm}{\includegraphics{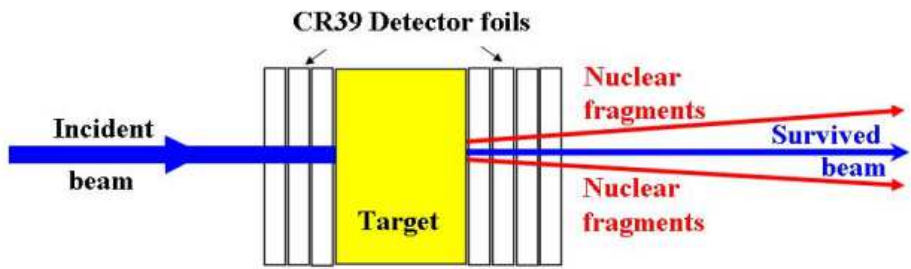}}\par}
\begin{quote}
\caption{\small Sketch of the target-detector configuration 
used for the exposures to different ion beams.} 
\label{fig:1}
\end{quote}
 \end{figure}

Stacks composed of several CR39 NTD's, of size $11.5 \times 11.5$ cm$^2$, and 
of different targets were exposed to 0.3, 1, 3, 5 and 10 A GeV Fe$^{26+}$, 1, 
3, 5 A GeV Si$^{14+}$ ions at BNL, 0.41 A GeV Fe$^{26+}$, 0.29 A 
GeV C$^{6+}$ ions at HIMAC. For these exposures we used the geometry 
sketched in Fig. \ref{fig:1}: three and four CR39 sheets, $\sim 0.7$ mm 
thick, were placed before and after the target, respectively. The exposures 
were done at normal incidence, with a density of $\sim 2000$ ions/cm$^2$. 
After exposures the CR39 foils were etched in 6N NaOH aqueous solution at 
70 $^\circ$C for 30 h (in two steps 15h+15h) in a thermostatic water bath 
with constant stirring of the solution. After etching, the beam ions and their 
fragments manifest in the CR39 NTD's as etch pit cones on both sides of each 
detector foil.

The base areas of the etch-pit cones (``tracks''), their eccentricity and 
central brightness were measured with an automatic image analyzer system 
\cite{17} which also provides their absolute coordinates. A tracking procedure 
was used to reconstruct the path of beam ions through the front faces of  
the detector upstream (with respect to the target) foils; a similar 
tracking procedure was performed through 
the three measured front faces of downstream CR39 detectors. The average 
track base area was computed for each reconstructed ion path by requiring 
the existence of signals in at least two out of three sheets of the detectors.
In Fig. \ref{fig:2}a,b the average base area distributions for 1 A GeV 
Si$^{14+}$ and 1 A GeV Fe$^{26+}$ beam ions and their fragments 
after the CH$_2$ targets are shown.

\section{Total fragmentation cross sections} 

The numbers of incident and survived beam ions were determined considering 
the mean area distributions of the beam peaks before and after the target 
and evaluating the integral of the gaussian fit of the beam peaks.

\begin{figure}[ht]
 \centering
{\centering\resizebox*{!}{6.5cm}{\includegraphics{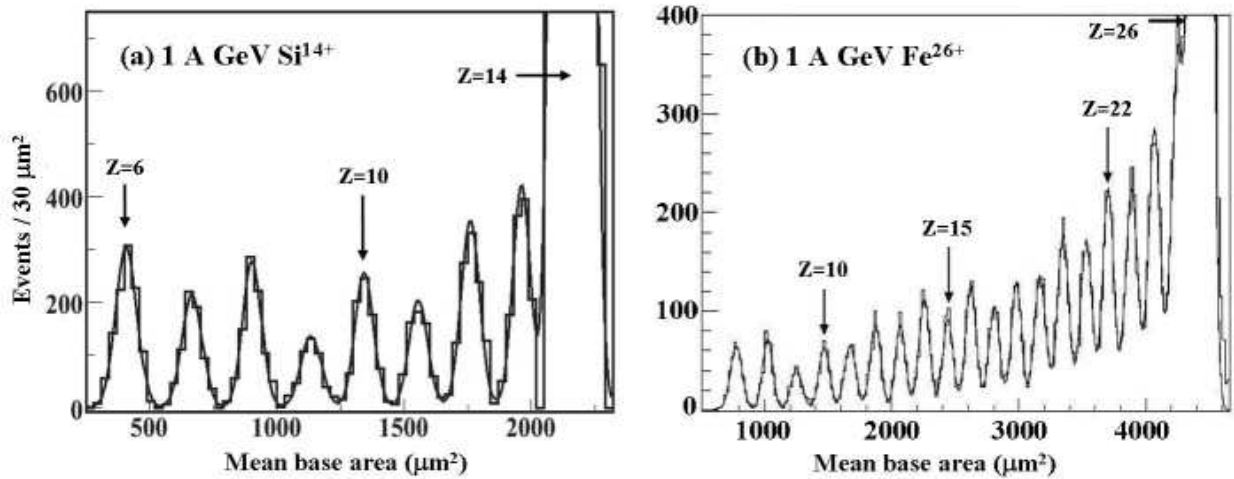}}\par}
\begin{quote}
\caption{\small Distributions of the average base areas for tracks present
in at least 2 out of 3 measured CR39 sheets located after the CH$_2$ 
target. The data concern (a) 1 A GeV Si$^{14+}$ and (b) 1 A GeV Fe$^{26+}$ 
ions. Each peak has a gaussian shape with $\sigma \sim 0.2e$. Notice that 
the peaks with $Z$ even are generally higher 
than the close by peaks with $Z$ odd.} 
\label{fig:2}
\end{quote}
\end{figure}

The total charge changing cross sections were determined with the survival 
fraction of ions using the following relation
 \begin{equation}
 \sigma_{tot} = \frac {A_T \ln (N_{in} / N_{out})}{\rho~ t ~N_{Av}}
\end{equation}
where $A_T$ is the nuclear mass of the target (average nuclear mass in case 
of polymers: $A_{CH2} = 4.7,~ A_{CR39} = 7.4$); $N_{in}$ and $N_{out}$ are the 
numbers of incident ions before and after the target, respectively; $\rho$ 
(g/cm$^3$) is the target density; $t$ (cm) is the thickness of the target and 
$N_{Av}$ is Avogadro number.\par

Systematic uncertainties in $\sigma_{tot}$ were estimated to be smaller 
than $10\%$: contributions arise from the measurements of the density 
and thickness of the targets, from the separation of the beam peak from the 
$\Delta Z = Z_{fragment} - Z_{beam} = -1$ fragments (Fig. \ref{fig:2}), from 
fragmentation in the 
CR39 foils and from the tracking procedure.  \par

\begin{figure}[ht]
\centering
 {\centering\resizebox*{!}{5.8 cm}{\includegraphics{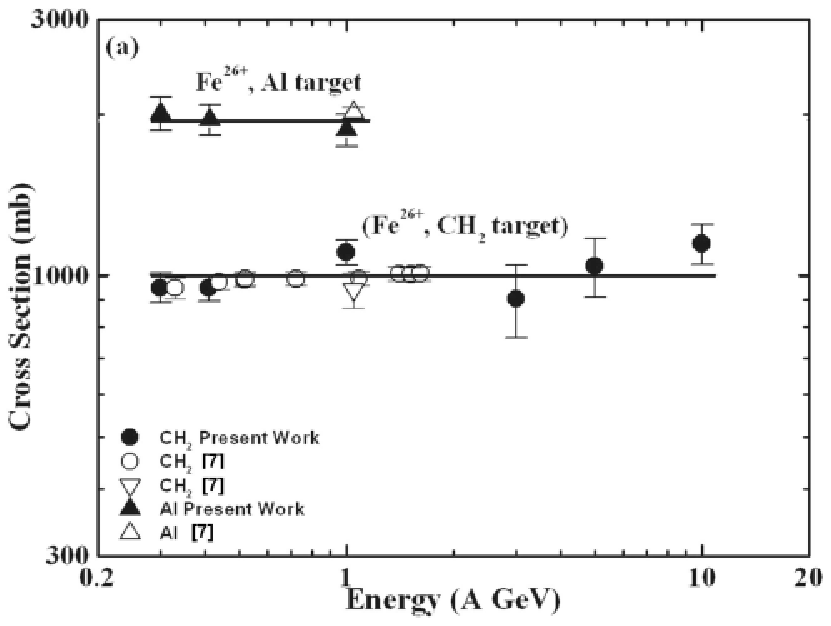}}}
 {\centering\resizebox*{!}{5.8 cm}{\includegraphics{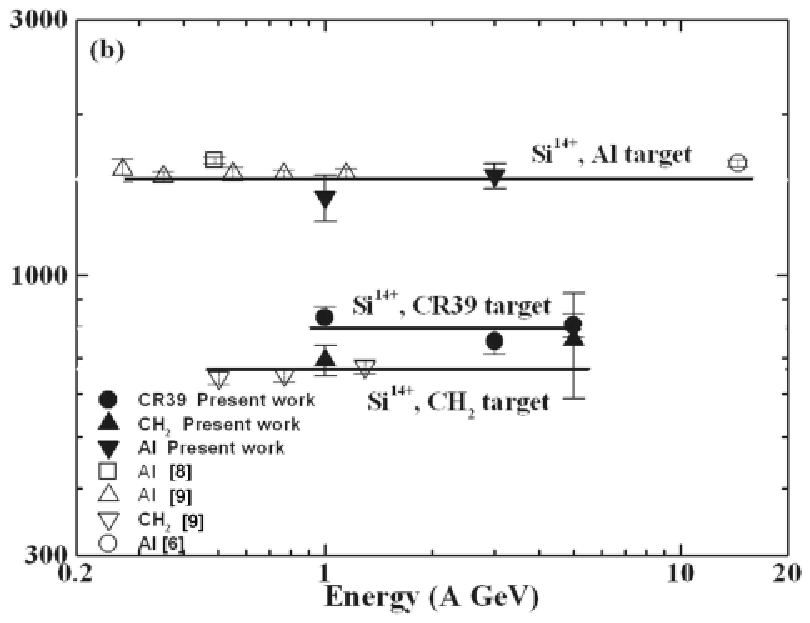}}\par}
\begin{quote}
\caption{\small Total fragmentation cross sections for (a) Fe ions of 
  different energies in CH$_2$ and Al targets and (b) for Si ions 
in CH$_2$, CR39 and Al targets. For 
comparison the measured cross sections from refs. \cite{6, 21, 23, 24} are 
also shown, together with the predictions from Eq. \ref{eq:2}.} 
\label{fig:3}
\end{quote}
 \end{figure}

The measured total charge changing cross sections are given in the $4^{th}$ 
column of Table \ref{table:1}. Fig. \ref{fig:3}a shows the total cross 
sections of Fe$^{26+}$ projectiles at various beam energies on the CH$_2$ 
and Al targets. Our results for 
Si$^{14+}$ and C$^{6+}$ projectiles are given in Table \ref{table:2} and are 
plotted vs energy in Fig. \ref{fig:3}b. 

The total cross sections are almost energy independent, in agreement 
with the data from other authors \cite{6, 21, 23, 24}.

\begin{table}
\begin{center}
{\small
\begin{tabular}
{|c|c|c|c|}\hline
{\bf Energy } & {\bf Target} & {\bf {\boldmath A$_T$}} & {\bf {\boldmath $\sigma_{tot}$} (mb)} \\
 {\bf (A GeV)} & & & \\ \hline 
 10 & CH$_2 $& 4.7 & 1147 $\pm$ 97 \\ \hline
 10 & CR39 & 7.4 & 1105 $\pm$ 360\\ \hline
 5 & CH$_2 $ & 4.7 & 1041 $\pm$ 130 \\ \hline
 5 & CR39 & 7.4 & 1170 $\pm$ 470 \\ \hline
 3 & CH$_2 $ & 4.7 & 904 $\pm$ 140 \\ \hline
 3 & CR39 & 7.4 & 1166 $\pm$ 67 \\ \hline
 1 & CH$_2 $ & 4.7 & 1105 $\pm$ 60\\ \hline
 1 & CR39 & 7.4 & 1113 $\pm$ 176 \\ \hline
1 & Al & 27 & 1870 $\pm$ 131 \\ \hline
 0.41 & CH$_2 $ & 4.7 & 948 $\pm$ 54 \\ \hline
 0.41 & CR39 & 7.4 & 1285 $\pm$ 245 \\ \hline
 0.41 & Al & 27 & 1950 $\pm$ 126 \\ \hline
 0.30 & CH$_2 $ & 4.7 & 949 $\pm$ 61 \\ \hline
 0.30 & CR39 & 7.4 & 1174 $\pm$ 192 \\ \hline
 0.30 & Al & 27 & 2008 $\pm$ 144 \\ \hline
\end{tabular}
}
\end {center}
\caption {Measured total fragmentation cross sections, with statistical 
standard deviations, for Fe$^{26+}$ ions of different energies (col. 1) on 
different targets (col. 2).} 
\label{table:1}
\end{table}

Various theoretical models/formulae for the total fragmentation cross sections 
were proposed and fitted to the experimental data 
with different geometrical radii and overlapping parameters \cite{8}. In 
Fig. \ref{fig:3} our data are compared with the semi-empirical 
formula \cite{14} for nuclear cross sections (solid lines)

\begin{equation}
\sigma_{tot} = \pi r_0^2~ (A_P^{1/3}  + A_T^{1/3} -b_0)^2
\label{eq:2}
\end{equation}
where $r_0 = 1.31$ fm, $b_0 = 1.0$, $A_P$ and $A_T$ are the projectile and 
target mass numbers, respectively. Various authors used different values 
for the overlap parameter $b_0$ within the interval $0.74 \div 1.3$ 
[5-10]. 

\begin{figure}[ht]
\centering
 {\centering\resizebox*{!}{6.0 cm}{\includegraphics{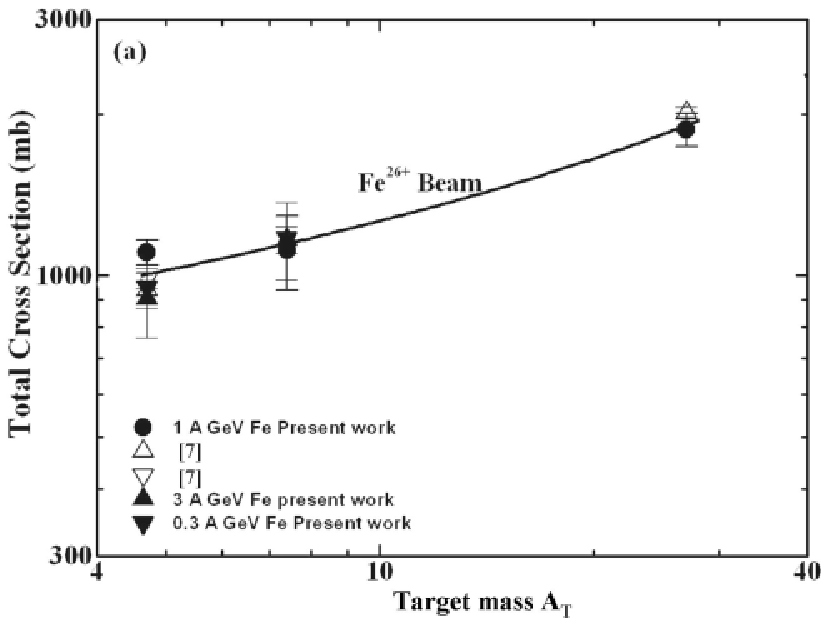}}}
 {\centering\resizebox*{!}{6.0 cm}{\includegraphics{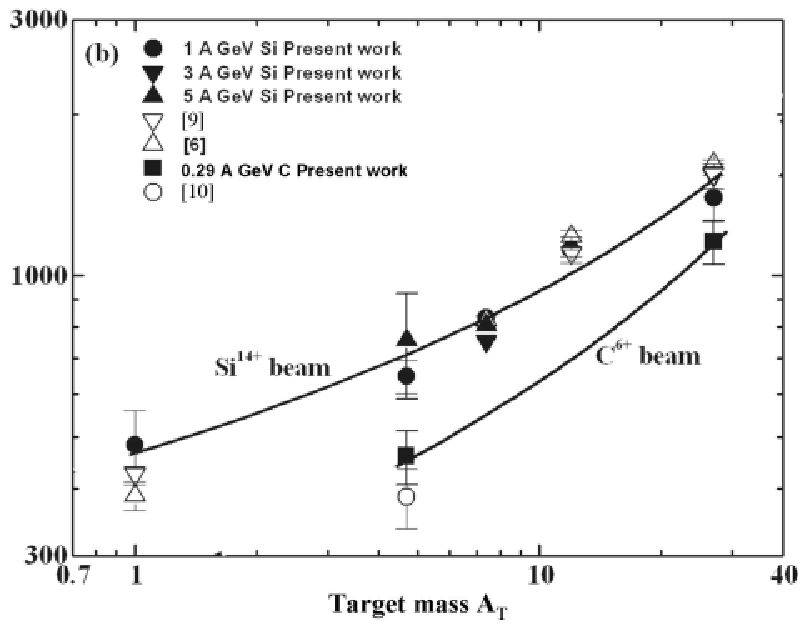}}\par}
\begin{quote}
\caption{\small Dependence of the total fragmentation cross 
sections on the target mass (a) for Fe ions and (b) for Si and C ions. For 
comparison the measured cross sections from refs. \cite{6, 21, 24, 9} are also 
shown. The solid lines are from Eq. \ref{eq:2} corrected by the $\sigma_{EMD}$
term.} 
\label{fig:4}
\end{quote}
 \end{figure}

Figs. \ref{fig:4}a,b show the total fragmentation cross sections vs target 
mass number $A_T$ for Fe$^{26+}$, Si$^{14+}$ and C$^{6+}$ beams of various 
energies. The solid lines are the predictions of Eq. \ref{eq:2}, to which 
we added the electromagnetic dissociation 
contribution, $\sigma_{EMD} = \alpha Z_T^{\delta}$, with 
$\alpha = 1.57$ fm$^2$ and $\delta = 
1.9$ [last ref. of \cite{11}]. The total fragmentation cross 
sections increase with increasing target mass number. Part of the 
increase is due to the effect of electromagnetic dissociation. \par

The data from other authors \cite{6, 21, 24, 9} are plotted for comparison
and show good agreement with our data, within the experimental uncertainties.

\begin{table}
\begin{center}
{\small
\begin{tabular}
{|c|c|c||c|c|c|}\hline
\multicolumn{3}{|c||}{\bf {\boldmath Si$^{14+}$} ions} & 
\multicolumn{3}{c|}{\bf {\boldmath C$^{6+}$} ions} \\ \hline
{\bf Energy} & {\bf Target} & {\bf {\boldmath $\sigma_{tot}$} (mb)} &
{\bf Energy} & {\bf Target} & {\bf {\boldmath $\sigma_{tot}$} (mb)} \\
{\bf (A GeV)} & & & {\bf (A GeV)} & &  \\ \hline
5 & CH$_2$ & 757 $\pm$ 168 & 0.29 & CH$_2$ & 460 $\pm$ 53 \\ \hline
3 & Al     & 1533 $\pm$ 133 & 0.29 & CR39 & 513 $\pm$ 52 \\ \hline
1 & CR39 & 1113 $\pm$ 176 & 0.29 & Al & 1155 $\pm$ 108 \\ \hline
1 & H & 483 $\pm$ 76 & & &  \\ \hline 
1 & CH$_2$ & 694 $\pm$ 70 & & & \\ \hline
1 & C & 1117 $\pm$ 62 & & & \\ \hline
1 & Al & 1397 $\pm$ 138 & & & \\ \hline

\end{tabular}
}
\end{center}
\caption {Measured total fragmentation cross sections $\sigma_{tot}$ for 
Si$^{14+}$ ions of different energies (col.1) on different targets 
(col. 2) and for 0.29 A GeV C$^{6+}$ ions on different targets 
(col. 5). Errors are statistical standard deviations.} 
\label{table:2}
\end{table}

\section{Partial fragmentation charge changing cross sections}
If the thickness of the target is small compared to the mean free path of 
the fragments in that material, the partial fragmentation cross sections 
can be calculated using the simple relation
\begin{equation}
\sigma(Z_i, Z_f) \simeq \frac{1}{Kt} \frac{N_f}{N_i}
\end{equation}
where  $\sigma (Z_i, Z_f)$ is the partial fragmentation cross section of an 
ion $Z_i$  into the fragment $Z_f$, $K$ is the number of target nuclei 
per cm$^3$, $t$ is the thickness of the target, $N_i$  is the number of 
survived ions after the target and $N_f$ is the number of fragments produced 
with charge $Z_f$. This expression may be valid also for a thick 
target, assuming that the number of fragments before the target is zero. \par

For the Fe ions, we observed that fragments are present even before the 
targets. In this case the partial charge change cross 
sections have been computed via the relation

\begin{equation}
\sigma_{\Delta Z} = \frac{1}{Kt} \left( \frac{N^f_{out}}{N^p_s} - \frac{N^f_{in}}{N^p_{in}} \right)
\end{equation}
where $N_{in}^{f}$ and $N_{out}^f$ are the numbers of fragments of each 
charge before and after the target, and $N_{in}^{p}$ and $N_{s}^p$ 
are the numbers of incident and survived projectile ions. \par
The distributions, after the CH$_2$ targets, of the fragments for 1 A GeV 
Si$^{14+}$ and 1 A GeV 
Fe$^{26+}$ ions are shown in Figs. \ref{fig:2}a,b. The relative partial 
fragmentation cross sections for 
$\Delta Z = -1, -2, -3,~..,~ -18$ are given in Table \ref{table:3}. The quoted 
errors are statistical standard deviations; systematic uncertainties are 
estimated to be about $10\%$. A clear odd-even effect is visible in Fig. 
\ref{fig:2}: the cross sections for the $Z-$even fragments are generally 
larger than those for the $Z-$odd fragments close by.

\begin{table}
\begin{center}
{\small
\begin{tabular}
{|c|c|c|}\hline
{\bf {\boldmath $\Delta Z$} } & {\bf 1 A GeV {\boldmath Fe$^{26+}$}} & 
{\bf 1 A GeV {\boldmath Si$^{14+}$}} \\ \hline
-1 & - & 293 $\pm$ 18 \\ \hline
-2 & 338 $\pm$ 11 & 177 $\pm$ 12 \\ \hline
-3 & 285 $\pm$ 11 & 123 $\pm$ 11 \\ \hline
-4 & 252 $\pm$ 10 & 122 $\pm$ 11 \\ \hline
-5 & 249 $\pm$ 10 & 62 $\pm$ 8 \\ \hline
-6 & 197 $\pm$ 9  & 117 $\pm$ 11 \\ \hline
-7 & 168 $\pm$ 8  & 83 $\pm$ 9 \\ \hline
-8 & 132 $\pm$ 7  & 90 $\pm$ 10 \\ \hline
-9 & 175 $\pm$ 8  & \\ \hline
-10 & 107 $\pm$ 7 & \\ \hline
-11 & 152 $\pm$ 6 & \\ \hline
-12 & 105 $\pm$ 8 & \\ \hline
-13 & 103 $\pm$ 6 & \\ \hline
-14 & 81 $\pm$ 6 & \\ \hline
-15 & 80 $\pm$ 6 & \\ \hline
-16 & 50 $\pm$ 4 & \\ \hline
-17 & 76 $\pm$ 5 & \\ \hline
-18 & 86 $\pm$ 6 & \\ \hline
\end{tabular}
}
\end {center}
\caption {The measured partial fragmentation charge changing cross sections 
for 1 AGeV Si$^{14+}$ and Fe$^{26+}$ ions on the CH$_2$ targets. The errors 
are statistical standard deviations. A systematic uncertainty of about $10\%$
should be added.} 
\label{table:3}
\end{table}

\section{Conclusions}
The total fragmentation cross sections for $^{56}$Fe, $^{28}$Si and $^{12}$C 
ion beams of $0.3 \div 10$ A GeV energies on polyethylene, CR39 
and aluminum targets were measured using CR39 NTD's \cite{19}. 

The total cross sections for all 
the targets and energies used in the present work do not show any observable energy 
dependence. There is a dependence on target mass; the
highest cross sections are observed for Al targets and this is mainly due to 
the contribution of electromagnetic dissociation. The present data of 
total fragmentation 
cross sections are in agreement with similar experimental data in the 
literature [4-10]. 

The presence of well separated fragment peaks, see Fig. \ref{fig:2}, allowed
the determination of the partial fragmentation cross sections. On the 
average the partial cross sections decrease as the 
charge change $\Delta Z$ increases. The data in Fig. \ref{fig:2} and the 
partial cross sections in Table \ref{table:3} indicate a clear $Z$ odd-even 
effect. \par

The measured cross section data indicate that passive NTD's, specifically 
CR39, can be used effectively for studies of the total and partial charge 
changing cross sections, also in comparison with active detectors. 

\section*{Acknowledgments} 

This work was in part financed by the MIUR PRIN 2004 Program (ex $40\%$), Prot.
2004021217.

We thank the technical staff of BNL and HIMAC for their kind cooperation 
during the beam exposures. We acknowledge the contribution of our technical 
staff, in particular of A. Casoni, M. Errico, R. Giacomelli, G. Grandi 
and C. Valieri. We thank INFN and ICTP for providing fellowships and grants 
to non-Italian citizens.

\bibliographystyle{plain}

\end{document}